\documentclass[pra,aps,amsmath, preprint, amssymb]{revtex4}
\usepackage{graphicx}
\usepackage{amsmath}
\usepackage{color}
\usepackage{float}

\begin{document}

\title{The role of substrates and environment in piezoresponse force microscopy: A case study with regular glass slides}

\author{Shilpa Sanwlani}
\author{Mohammad Balal}
\author{Shubhra Jyotsna}

\author{Goutam Sheet}
\email{sgoutam@gmail.com}
\affiliation{Department of Physical Sciences,  
Indian Institute of Science Education and Research Mohali,
Punjab, India, PIN: 140306}

\begin{abstract}
Piezoresponse force microscopy (PFM) is a powerful tool for probing nanometer-scale ferroelectric and piezoelectric properties. Hysteretic switching of the phase and  amplitude of the PFM response are believed to be the hallmark of ferroelectric and piezoelectric behavior respectively. However, the application of PFM is limited by the fact that similar hysteretic effects may also arise from mechanisms not related to ferroelectricity or piezoelectricity. In this paper we report our studies on regular glass slides that show ferroelectric-like signal without being ferroelectric and frequently used as a substrate in PFM experiments. We demonstrate how the substrates and other environmental factors like relative humidity and experimental conditions may influence the PFM results on novel materials.

Keywords: Piezoresponse force microscopy(PFM); Ferroelectrics; Hysteresis; Piezoelectrics.
\end{abstract}

\maketitle

Piezoresponse force microscopy (PFM) is one of the most powerful techniques used for imaging and identifying domains in a ferroelectric and piezoelectric materials  \cite{Roger, Lines, Kalinin, Shvartsman, Shvartsman1, Kholkin}. This is done by bringing a sharp electrically conductive tip in contact with the surface of the material and applying an external bias to the conductive tip.  An electromechanical interaction between the cantilever tip and the sample determines the sample response to an applied sweeping dc electric field. As the electric field sweeps, an ac signal rides on the sweeping field which is used as an excitation signal in order to track the sample characteristics \cite{Soergel, Kalinin1}. A phase switch of the response and hyteresis in phase vs. dc voltage plot is known as an evidence of local polarization switching indicating ferroelctricity, whereas, hysteresis in amplitude with field indicates piezoelectric behaviour \cite{Kalinin1}.

It is seen that in some of the materials, which do not have ferroelectric ordering may show strikingly similar hysteretic switching behavior under a PFM \cite{Jagmeet}. The origin of hysteresis in such materials is poorly understood. Such observations are usually attributed to the local electrostatic charge accumulation or local electrochemical effects. The role of the electrostatic effects are usually mitigated by employing the switching spectroscopy PFM (SS-PFM), protocol pioneered by Jesse $et. al$, in which the dc-bias is applied in a sequence of electric pulses and the measurements are done in the ``off"-states of those pulses \cite{Jesse, Jesse1, Jesse2}. Since the electrostatic charges are expected to relax much faster than the ferroelectric polarization, this technique removes the electrostatic components from the measured data to a large extent. While the electrochemical effects cannot be entirely removed, whether electrochemistry contributes to the observed switching or not can be investigated by performing topographic imaging after performing PFM spectroscopy. In case of electro-chemical reactions taking place under the tip additional nano-structures are known to grow as an end product of the said electrochemical reaction that can be imaged by regular non-contact topographic imaging \cite{Campbell, Wang, Wang1, Asmah, Dagata, Tello}. In fact, recently the tip-induced electrochemistry has been exploited in a new charaterization technique namely  Electrochemical Strain Microscopy (ESM) \cite{Roger} which has been applied to enormous number of non piezoelectric but electrochemically active materials such as in biological molecules \cite{Kalinin2, Kalinin3}, corrosion and in medical science \cite{Roger, Nazri, Abraham, Hayre, Bansal}. It was observed that these substances exhibit piezoelectric and ferroelectric- like behaviour on interaction with the electrically conductive tip. In PFM spectroscopy on such materials spectral features such as hysteretic phase switching along with the hysteresis in amplitude (butterfly loops) can be observed as well. In some cases it was also shown that such ferro-electric-like signal might be associated with hysteretic switching behaviour of the substrate or the base material on which the samples are mounted \cite{Jagmeet}.

\begin{figure}
	\centering
		\includegraphics[width=0.80\textwidth]{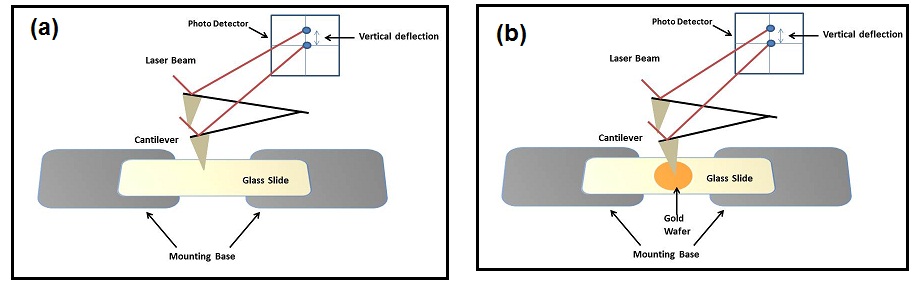}
	\caption{(a) Schematic diagram of piezoresponse force microscopy (PFM) on a regular glass slide. (b) Schematic diagram of PFM on a gold (Au) wafer mounted on the glass slide.}
	\label{Figure 1}
\end{figure}

One of the popular base materials on which PFM samples are usually mounted are the regular glass slides. The primary chemical component of such glass slides is amorphous silica. However, most of them are known to contain significant amount of rare-earth impurities like sodium \cite{Zacharaisen, Mehrer}. Sodium containing glass slides are known to be electrochemically active \cite{Roger}. However, from structural point of view glass slides are not expected to be ferroelectric or piezo-electric. This makes the slides a natural choice as substrates for PFM measurements \cite{Suenne}.  Recently it has been shown that hysteretic effects in sodium-glass slides might originate from electrochemical reactions taking place on the surface of the glass slide \cite{Roger}. In ref. \cite{Roger} it was shown that during the PFM spectroscopic measurements on glass slides using conductive tips, nano-meter size topographic structures grew that could be imaged in the non-contact AFM mode after the spectroscopic measurements \cite{Xie}. This observation is a proof of the electro-chemical processes in a small region on the sample surface underneath the tip \cite{Balke, Balke1}. However, on other glass slides with different components, signature of electrochemical reaction are not found by topographic imaging after the spectroscopic experiments. Therefore, the hysteresis effects observed on low sodium glass slides cannot be simply attributed to electrochemically induced switching. Other experimental and environmental factors need to be considered. In addition, our observation has been that every material that is mounted on a glass slide shows hysteretic effects under a piezo-response force microscope. Therefore, the role of substrates in PFM experiments should be investigated. 

In this paper we report our measurements on a wide variety of samples that do not show hysteresis in PFM response normally, but do show hysteretic phase and amplitude switching when mounted on glass slides. We also performed spectroscopic measurements on a diverse variety of samples including paper which is known to be a good dielectric and a small piece of thin Magnesium Oxide (MgO) and pure metal, gold. We do not observe hysteresis on these systems when they are directly mounted on the PFM sample holder. However, we could observe the hysteretic effects in these materials when mounted on the glass slide. We discuss these results in the light of the possible role of the grounding path and other environmental factors in PFM experiments.

	 \begin{figure}
		 \centering
			 \includegraphics[width=0.80\textwidth]{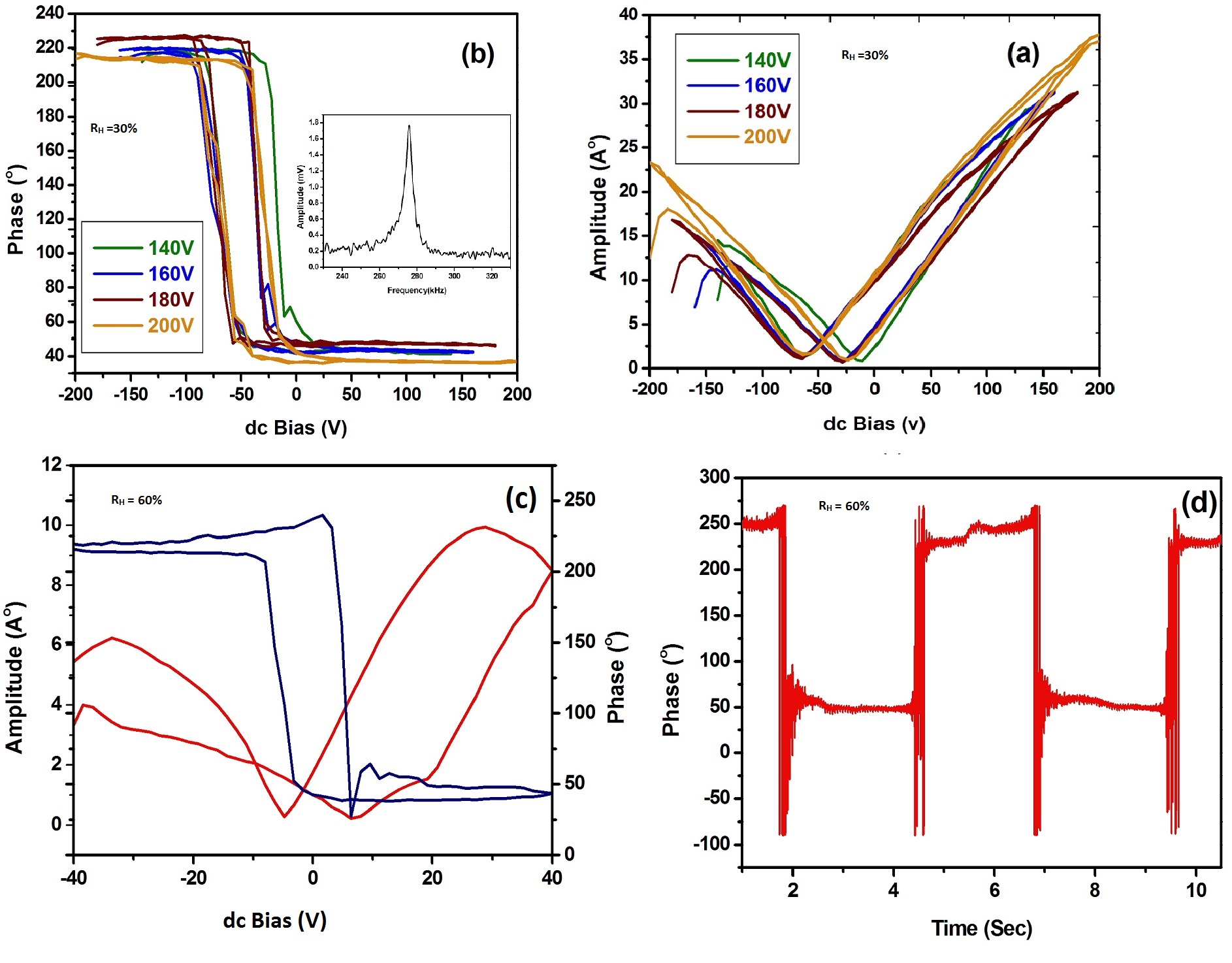}
		 \caption{(a) PFM phase hysteresis on glass slide upto 200V with inset showing that in-contact
resonance frequency of the cantilever tip varied between 270 kHz to 280 kHz. (b) Butterfly loops
on glass slide upto 200V.(c) PFM hysteresis and butterfly loops observed in glass slide at 160 Volts
at relative humidity 60{$\%$}. (d) Typical phase-switching in time domain at relative humidity 60{$\%$}.
}
	 \label{Figure 2}
 \end{figure}

The complete set of experiments was performed with a commercial atomic force microscope (AFM) from Asylum Research (model:MFP3D) \cite{Zacharaisen, Rodriguez}. For this study, we used Dual AC Resonance Tracking(DART) mode of PFM. This mode enables us to study the spectroscopic properties of the material along with identifying domains in a ferroelectric/piezoelectric material. A sharp conductive cantilever tip of silicon(Si) coated with Ti/Ir(5/20) with radius of 28+/-10 nm and spring constant 2N/m was used. The natural resonance frequency of the cantilever tip in air was 70kHz. A regular microscopic glass slide of 2 mm thickness was taken as the substrate material. The tuning curve is shown in the inset of FIG. 2(a). The in-contact resonance frequency of cantilever tip varied between 270 kHz to 280 kHz.

FIG. 1 gives the schematic description of the experimental setup. In FIG. 1, the grey part shows the regular mounting base for mounting substrate and the sample in the AFM. On the grey mounting base, glass is shown mounted in faded yellow color. The electrically sharp conductive cantilever tip is shown with brown color, interacting with the glass slide surface. The laser beam hits the cantilever and reflects back to the photo-detector in the AFM. The deflection by the cantilever tip is noted by the photo-detector. The vertical deflection is indicated using dark blue color on the photo-detector. In FIG. 1(b), we showed a gold (Au) wafer mounted on the glass slide. All the samples mentioned here were mounted in the same way.

As mentioned before, in order to mitigate the effect of local charging and local static electric field the measurements have been done in switching spectroscopy (SS-PFM) while we applied a sequence of dc bias voltage to the conductive cantilever tip. We used a high voltage amplifier to apply bias voltage upto 200 volts to the cantilever tip during the spectroscopic measurements. It was observed that for the lower voltage range, there were no ferroelectric/piezoelectric- like response. This set of experiments was performed by keeping relative humidity of the lab at 30{$\%$}. However, at higher bias voltages it showed significantly repeating loops of hysteresis with phase and hysteresis with amplitude, similar to that in piezoelectrics. In FIG. 2, it can be seen that hysteresis and butterfly loops are visible at higher voltages, starting from 140 volts. Below this voltage, we attempted to switch the conductive probe at several points on the surface of the glass slide down to 10 volts but we could not find any such signals. In FIG. 2(a), hysteresis with phase has been shown. From the visual inspection, on applying higher and positive dc bias to the cantilever tip, the phase angle was 40 degrees, while on applying negative bias of 150 volts, the phase switched at 220 degrees, signifying a clear phase switching. Similarly, in FIG. 2(b), we got two dips in  the butterfly loops, one on applying positive dc bias and second on applying negative dc bias. 
In addition, the relative humidity of the lab was varied from 30{$\%$} to 60{$\%$}. In the presence of higher humidity levels, the glass slides showed hysteretic features at lower bias voltages. As shown in FIG. 2(c),  the clear evidence of hysteretic effects on the glass slide were observed at maximum range of 40 volts when the relative humidity was kept at 60{$\%$}. FIG. 2(d) is a representation of graph showing switching of phase in time domain.

\begin{figure}
	\centering
\includegraphics[width=0.80\textwidth]{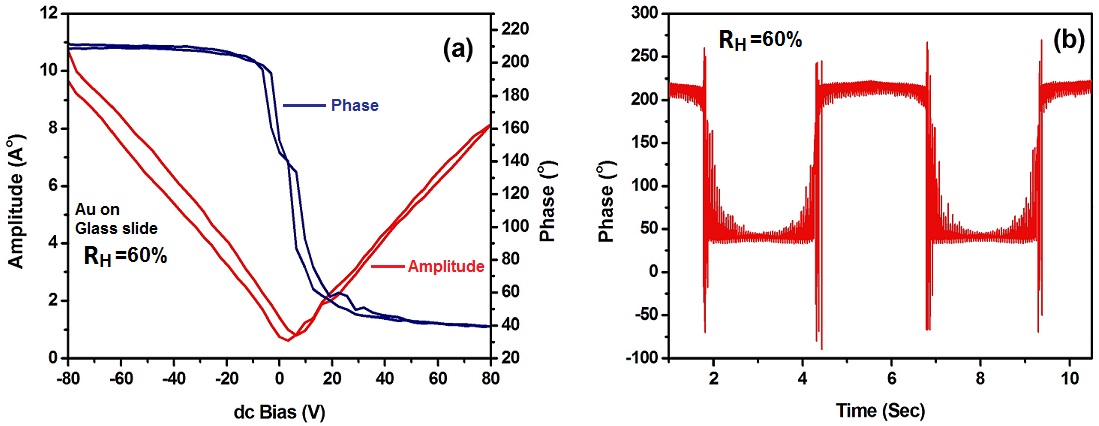}
	\caption{(a) PFM phase hysteresis and butterfly loop on gold (Au) mounted on glass slide at 80 volts at relative humidity 60{$\%$}. (b) Typical phase-switching in time domain at relative humidity 60{$\%$}.}
	\label{Figure 3}
\end{figure}

Hysteretic switching was observed on bare glass slides when mounted on the sample holder. In order to observe the effect of this switching on other samples mounted on a glass slide, we mounted a wide variety of non-ferroelectric/non-piezoelectric samples on the glass slide that include a gold disk, a single crystal of MgO and a piece of regular paper.We repeated the same set of experiments with a 0.125 inch dia circular disk of pure gold (Au) as  a sample mounted on the glass slide. Gold is known to be non-ferroelectric/non-piezoelectric material. However, when we performed PFM experiments on Au by sweeping the dc bias voltage upto 100V in the presence of high humidity of 60{$\%$}, we observed hysteretic phase and amplitude switching. When we mounted the gold (Au) disk directly on the AFM sample holder and not on a glass slide, as expected, it did not exhibit ferroelectric/piezoelectric-like spectroscopic features. The results on Au are presented in Fig. 3. 
We also performed switching spectroscopy on some known non-conducting and non-ferroelectric materials, like a small piece of regular paper and a thin crystal of MgO. We found that these trivial materials also exhibited ferroelectric/piezoelectric-like signals, like pure gold disk exhibited when placed on the glass slide. FIG. 4 gives the clear evidence of local hysteretic effects in paper piece and MgO wafer. It should be noted that the way the sample sees the ground is different when it is mounted on the AFM sample holder (which is directly connected to the ground) and on the glass slide. In case of glass slide the sample may find a ground path through thin natural layer of moisture on the glass slide. 
\begin{figure}
	\centering
		\includegraphics[width=0.80\textwidth]{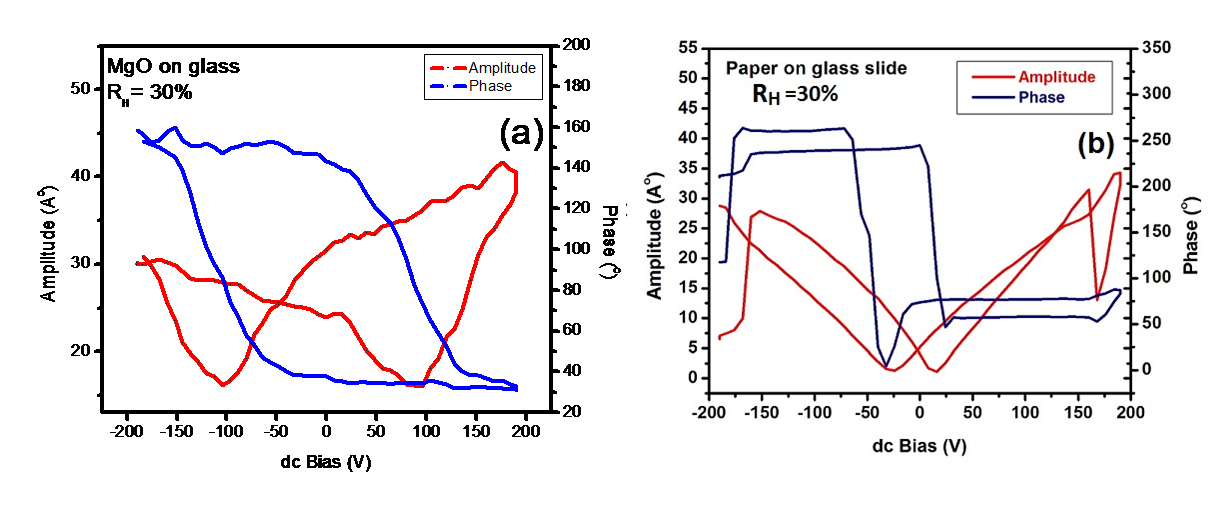}
	\caption{(a)PFM phase hysteresis and butterfly loop on MgO mounted on glass slide. (b) PFM phase hysteresis and butterfly loop on paper mounted on glass slide.}
	\label{Figure4}
\end{figure}

\begin{figure}
	\centering
		\includegraphics[width=0.50\textwidth]{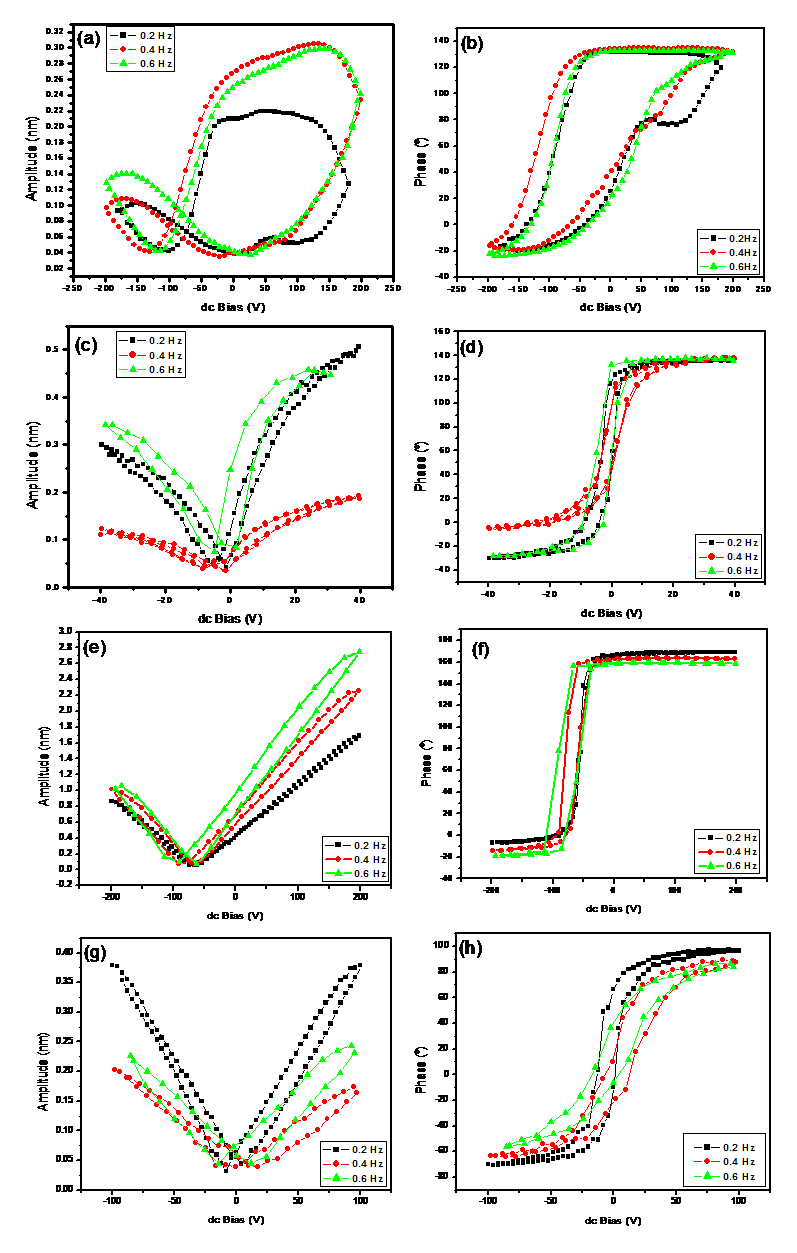}
	\caption{PFM spectroscopy measurements at different frequency sweep rates of 0.2Hz, 0.4Hz and 0.6Hz showing butterfly loops for (a) glass, (c) gold on glass substrate, (e) MgO on glass substrate and (g) paper on glass substrate; and hysteresis loops for (b) glass, (d) gold on glass substrate, (f) MgO on glass substrate and (h) paper on glass substrate}
	\label{Figure 5}
\end{figure}

It is important to investigate the time relaxation dynamics of the data in order to study the time dependence strain behavior of piezoelectric material or glass slide type sample. As shown in Figure 5 we have performed the measurement by changing the rate at which we apply the voltage pulses for switching spectroscopy experiments on bare glass, gold on glass substrate, MgO on glass and paper on glass at a high humidity level (above 45{$\%$}) where we observe the dip in phase-voltage curve. All the measurements were done, as for the data in the OFF state. From the Figure 5 we observe that all the four samples exhibit strong time dependence with longer periods resulting in higher strain responses. It is evident from the butterfly loops under different voltage sweep cycles of 0.2Hz, 0.4Hz and 0.6Hz. In consistence with the Chen et al.\cite{Chen} this behavior is attributed with the longer range of ionic redistribution of glass and thus larger induced dipoles. We have also done measurements for the several sweep cycles and averaged them to gain insight on the effect of increasing number of sweep cycles. The observation is, as the number of sweep cycles are increased, the butterfly and hysteresis loops exhibit longer periods in the amplitude and phase respectively as shown in Figure S1 in supplementary material . And this observation is consistent for all the four samples i.e. glass, gold on glass substrate, MgO on glass substrate and paper on glass substrate. All the relevant information extracted from these set of measurements have been added to the supplementary material.

In conclusion, regular glass slides showed hysteretic switching on performing switching sprectroscopy PFM (SSPFM) when placed on the sample holder in AFM. Switching features were also observed on the surface of some non-ferroelectric/non-piezoelectric materials like pure gold, single crystal of MgO and a regular paper when mounted on the glass slide. This might be attributed to the difference in the way the samples find the ground path when they are mounted on the glass slides compared to when they are directly mounted on the AFM sample holder that is electrically shorted to the ground of the high-voltage amplifier used for PFM. This is supported by the fact that the switching behaviour depends on the level of humidity in the atmosphere -- the samples mounted on the insulating glass slides get connected to the electrical ground of the equipment through natural layers of moisture present on the glass slides. Therefore, it is rational to conclude that regular glass slides are not suitable for application as substrates in PFM measurements and the materials that were earlier reported to show ferroelectricity/piezoelectricity by PFM experiments and were mounted on glass slides should be revisited. As glass and glass-like samples discussed in the paper are non-piezoelectric, a characteristic behavior cannot be shown in the butterfly loop like for a piezoelectric material saturation is expected in the butterfly loops. At this point, we believe more experiments need to be done to evaluate butterfly loops for non-piezoelectric materials.

We thank L. Aggarwal for her help. GS acknowledges the research grant of Ramanujan Fellowship from Department of Science and Technology (DST), India for partial financial help.

\end{document}